\documentstyle[preprint,aps]{revtex}
\begin{document}

\draft
\title
{\bf Effects of electron correlation on the photocurrent in
quantum dot infrared photodetectors$^*$}

\author{Yia-Chung Chang and David M.-T. Kuo}

\address
{Department of Physics and Materials Research Laboratory\\
University of Illinois at Urbana-Champaign, Urbana, Illinois 61801}
\date{\today}
\maketitle

\begin{abstract}
The effect of electron correlation on the photocurrent of self-assembled
InAs/InGaAs quantum dot infrared photo-detector (QDIPs) is
studied. It is found that Coulomb interaction and level mixing in
the many-body open system lead to double peaks associated with the
intra-band transitions involving two lowest levels of the quantum
dot. Furthermore, the photocurrent is a nonlinear function of the
steady-state carrier density and it displays a plateau due to
Coulomb blockade.
\end{abstract}

\mbox{}\\
\vspace{2in} \mbox{}\\
${}^*$This work was supported by a subcontract from the University
of Southern California under the MURI program, AFSOR, Contract No.
F49620-98-1-0474.
\newpage

Recently, many efforts have been devoted to the understanding of
transport properties in quantum dot photo-detectors (QDIPs)[1,2].
The advantage of the QDIP over quantum well photo-detectors (QWIPs)
is that light can be directly coupled to the electrons in the
normal incidence geometry due to the effect of QD confinement in
directions perpendicular to the growth axis and the dark current
is smaller for the same detection wave length considered[3].
Other significant features that are unique to QDs include the
Coulomb blockade effect[2] and phonon bottleneck[4].

    Due to the localized nature of electrons in QDs, it is
essential to take into account the effects of Coulomb blockade in
the analysis of photoresponse of QDIPs, which in general can be
ignored in QWIPs. For the non-equilibrium system considered here,
it is convenient to use the Keldysh Green function to calculate
the transport and optical properties while including the electron
correlation. This technique has been used extensively in the study
of nonlinear transport properties of quantum systems[5,6].

We solve the Anderson Hamiltonian of a two level system coupled
with leads in the presence of an electromagnetic radiation with
frequency $\omega$. We adopt the approach given by Jouho et
al[5,6] and extend it to the present case with asymmetric
tunneling rates. We find that the time averaged tunneling current
is given by

\begin{equation}
\langle J(t)_j \rangle =  ( \Gamma_j^R- \Gamma_j^L)
\frac{e\langle N_j(t) \rangle }{2} - e \int
\frac{d\epsilon}{\pi} [ \Gamma^L_j f_L- \Gamma^R_j f_R ]
Im G^r_j(\epsilon, \omega)
\end{equation}
where  $\langle N_j(t) \rangle $ is the electron occupation number
at QD, $f_L=f(\epsilon-\mu_L)$ and $f_R=f(\epsilon-\mu_R )$ are
the Fermi distribution function of the left lead and right lead,
respectively. The chemical potential difference between these two
leads is related to the applied bias ($V_a$) via $\mu_L -\mu_R =
eV_a $. $\Gamma^L_j $ and $\Gamma^R_j$ denote the tunneling rates
from the QD to the left and right leads, respectively, for
electrons in level $j$. $G^r_j(\epsilon,\omega)$ is the
retarded Green function for an electron in level j of the QD.
The first term in Eq. (1) for the excited state $(j = 2)$ provides the
photo-induced tunneling current which exists only when $\Gamma^R_2
\neq \Gamma^L_2$, a condition that can occur in a system
with asymmetric potential.

Since the incident radiation considered in QDIP application is
usually very weak, we can ignore the renormalization of
the retarded Green's function due to electron-photon interaction
. Thus, we have (within the Hartree-Fock approximation)

\[
G^r_{1}(\epsilon) = \frac{1-N_1}{\epsilon -
E_{1} + i\Gamma_{1}/2} +\frac{N_1}{\epsilon -
E_{1} - U_{11} + i\Gamma_ {1}/2},
\]
\[
G^r_{2}(\epsilon) = \frac{1-N_1}{\epsilon -
E_{2} + i\Gamma_{2}/2} +\frac{N_1}{\epsilon  -
E_{2} - U_{12} + i\Gamma_ {2}/2},
\]
where $U_{11}$ denotes the Coulomb interaction between two
electrons in level 1 and $U_{12}$ denotes that for one electron in level 1
and the other in level 2. To calculate the steady-state electron
occupation number for the QD, we solve the semiconductor Bloch
equations for the two-level system coupled to leads and the electromagnetic
radiation.[6] We obtain (for $N_2 \ll N_1$)

\[ \Gamma_1 N_1  =  -Im{\cal X} - \int \frac{d\epsilon}{\pi}
[\Gamma_1^{L} f_{L}(\epsilon) + \Gamma_1^{R} f_{R}(\epsilon)]
ImG^r_{1}(\epsilon), \]
\[ \Gamma_2 N_2  =  Im{\cal X} - \int
\frac{d\epsilon}{\pi} [\Gamma_2^{L} f_{L}(\epsilon) + \Gamma_2^{R}
f_{R}(\epsilon)] ImG^r_{2} (\epsilon),
\]
and
\[
{\cal X}(\omega) = 2\lambda^2 (N_2 -N_1) \{ \frac{1-N_1}{\omega_r
- \omega + i \Gamma/2}
 +\frac{N_1}{\omega_r+U_{12}-U_{11} -
\omega + i \Gamma/2} \},
\]
where $\omega_r \equiv E_2 - E_1$ is the resonant frequency, $\Gamma \equiv
\Gamma_1 + \Gamma_2$, and $\lambda$ is the momentum matrix element for
the inter-level optical transition.

Note that ${\cal X}(\omega)$ consists of two poles corresponding
to optical transitions from initial states at $E_1 $ and  ($
E_1+U_{11}$) to final states at $E_2$ and $(E_2 + U_{12})$,
respectively. The two transitions have different relative
strengths which depend on the averaged steady -state occupation
number in the ground state, $N_1$. The first term corresponds to the inter-level transition of
a single electron in the QD [which occurs with a relative
probability $(1-N_1)$], while the second term corresponds to the
inter-level transition of a second electron in the QD (which
occurs with a relative probability $N_1$) under the influence of
the first electron, which remains in the ground state at all
times. In the latter case, the Coulomb repulsion between the two
electrons give rise to an energy shift $U_{12} - U_{11}$. The fact
that we have a fractional occupancy $N_1$ in a single quantum dot
is attributed to the level mixing effect (coupling of the QD level
to the continuum states in the leads) in the many-body open
system.


The theory is applied to a realistic self-assembled quantum dot
(SAQD) device. We consider an InAs/InGaAs SAQD system with conical
shape. The SAQD is embedded in a slab of InGaAs with
finite width, $W$. The slab is then placed in contact with heavily
doped InGaAs to form an n-i-n structure for infrared detection.
Within the effective-mass model[3], the QD electron is described by
the equation
\[ [-\nabla \frac 1 {2m^*(\rho,z)} \nabla + V(\rho,z) - eFz]\psi(\rho,\phi,z)
= E \psi(\rho,\phi,z), \] ${m^*(\rho,z)}$ is the
position-dependent effective mass, which takes on values of $m^*_G
=0.067 m_e$ (for GaAs) and $m^*_I = 0.024 m_e$ (for InAs). The
potential $V(\rho,z)$ is equal to $0$ in the InGaAs barrier region
and $V_0$ inside the InAs QD region. The potential in the
depletion layers (which separate the slab from the leads) are
modelled by an electrostatic potential $V_d(z)$

\[ V_d(z) = \left\{ \begin{array}{ll}
\frac{-V_1}{D} (z+\frac W 2) & \mbox{for }\;  -(D+\frac W 2)<z<-\frac W 2\\
\frac{V_1}{D}(z-\frac W 2) & \mbox{for } \;  \frac W 2<z<D+\frac W
2.
\end{array}
\right. \]
For the purpose of constructing the approximate wave functions, we
place the system in a large cylindrical confining box with length
$L$ and radius $R$ ($R$ must be much larger than the radius of the
cone, $r_c$). We adopt $R=400$\AA, $D =  350$ \AA,
$V_1 = -0.205 eV$, and $ W = 300 $\AA  for all calculations.
We solve the eigen-functions of the effective-mass
Hamiltonian via the Ritz variational method. The wave functions
are expanded in a set of basis functions which are chosen to be
products of Bessel functions and sine waves
\[
\psi_{nlm}(\rho,\phi,z)= J_l(\alpha_n\rho) e^{il\phi} \sin [k_m
(z+\frac{L}{2})],
\]
where $k_m = m\pi/L$, m=1,2,3... Throughout the paper, the origin
of z is set at the middle of the confining box.
$J_l$ is the Bessel function of order $l$ ($l=0,1,2,...,$
etc.) and $\alpha_nR$ is the $n$-th zero of $J_l(x)$.
40 sine functions multiplied by 15 Bessel functions for each
angular function ($l=0$ or 1) are used to diagonalize the
Hamiltonian.

Fig. 1 shows the energy levels of the confined states with $l=0$
(solid line) and 1 (dotted line) as functions of height $h$ of the
QD with base radius fixed at $R_0=70$ \AA $\mbox{}$. The other
material parameters used here are: wetting layer thickness $d = 3$
\AA, the conduction-band offset $V_0=-0.4$ eV (this includes the
effect of hydrostatic strain due to the lattice mismatch between
InAs and In$_{0.2}$Ga$_{0.8}$As), and length of the confining box
$L$ = 600 \AA. At least two bound states for each angular function
($l=0,$ or 1) are found. For infra-red detector application, we
are seeking an intra-band transition (between the ground and first
excited state) at an energy around 0.125 eV, which occurs at $h =
50$ \AA, for $R_0 = 70$ \AA. The tunneling rates can be calculated
numerically via the stabilization method as described in Ref. 3.

We only consider the zero temperature and low bias case, where the
chemical potential at the left leads ($\mu_L$) is lower than
$E_2$, so that the average population in the exited state is
small. Fig. 2 shows the photocurrent as a function of frequency
for various applied voltages: solid line ($V_a = 0.11 V$), dotted
line ($V_a = 0.12 V$), and dashed line ($V_a = 0.13 V$). The
parameters used to obtain Fig. 2 are $E_1 = -139 meV$, $E_2 = -14
meV $, $U_{11} = 10.4 meV$, and $U_{21} = 7.2 meV$, which are all
calculated based on the effective-mass model. The Fermi level in
the source and drain region is assumed to $E_F = 15 meV$. The
broadening of the energy level $E_1$ including all tunneling
processes (dominated by the acoustic-phonon assisted tunneling in
this case) is assumed to be $\Gamma_1 = 0.01 meV$. The precise
value of $\Gamma_1$ is not important, since photocurrent is not
sensitive to $\Gamma_1$. For the excited state, the broadening
parameter is given by $\Gamma_2=\Gamma'_{2}+\Gamma^R_2$.
$\Gamma'_{2}$ is mainly due to radiative and non-radiative
recombination from interacting with phonons and defects. The
actual value depends on the sample quality and temperature. Here,
we assume $\Gamma'_{2}= 1 meV$. The other contribution due to the
direct tunneling is calculated via the stabilization method as
described in Ref. [3]. The values are found to be  $\Gamma^R_2 =
0.439 meV$, $ 0.545 meV$, and $ 0.651 meV$  for $V_a = 0.11,
0.12$, and $ 0.13 V$, respectively. The spectrum of photocurrent
consists of two peaks centered at frequencies $\omega = E_2 -
E_1$, and $\omega = E_2 - E_1 + U_{12} - U_{11}$. The relative
strength of these peaks are determined by the average occupation
number in the ground state $(N_1)$, which is bias-dependent. As
shown in the figure, the Coulomb interaction leads to a
double-peak photocurrent spectrum with energy separation related
to the intra-level and inter-level Coulomb energies ($U_{11}$ and
$U_{12}$).

For the QDIP characteristics, the photocurrent versus applied bias
($J-V$ curve) is also of interest[7-9]. Fig. 3 shows the
photocurrent as a function of bias for frequencies at $\omega_r$
(dotted line) and $\omega_r+ U_{12}-U_{11}$ (solid line). Using
Eq. (2) for the polarization, we can readily understand the
behavior of photocurrent. The behavior of the photocurrent is
determined by the prefactors $(1-N_1) N_1$ and $ N_1 N_1 $ for the
two poles at $\omega = \omega_r$ and $\omega = \omega_r + U_{12} -
U_{11}$, respectively. At very low bias, $N_1$ is small; thus, the
magnitude of the solid line is much weaker than that of the dotted
line. As the applied bias increases, the solid line displays a
plateau due to the effect of Coulomb blockade on $N_1$. When the
applied bias overcomes the charging effect, $N_1 \approx 0.5$ and
the solid line becomes almost identical to the dotted line.


In this study, we have used tunneling carriers as the source for
photocurrent, in contrast to the captured carriers typically used
in QWIPs. Due to the phonon bottleneck effect[4], it is predicted
that the capture rate of the electron by the QD will be low. This
could reduce the performance of QDIPs which use captured carriers
as a photocurrent source. Using tunneling carriers as the
photocurrent source will not have this drawback, and in this case
we find that the effect of electron
correlation leads to a double-peak spectrum and the photocurrent
is a highly nonlinear function of the carrier density in the
ground state $N_1$. Both of these effects must be taken into
account in the analysis of photoresponse of QDIPs.



{\bf Figure Captions}

Fig. 1. Energies of the bound states of a conical
InAs/$In_{0.2}Ga_{0.8}$As QD as functions of height $h$ of the QD
with base radius fixed at $R_0=70$ \AA $\mbox{}$. Solid lines: ($l
= 0$). Dotted lines: ($l = 1$).

Fig. 2. Photocurrent as a function of frequency for different
applied voltages: $V_a = 0.11 V$ (solid line), $V_a = 0.12 V$
(dotted line) and $V_a = 0.13 V$ (dashed line).

Fig. 3. Photocurrent as a function of bias for incident frequencies
at $\omega = E_2 - E_1$ (dotted line) and $\omega= E_2 - E_1 + U_{12}
-U_{11}$ (solid line).

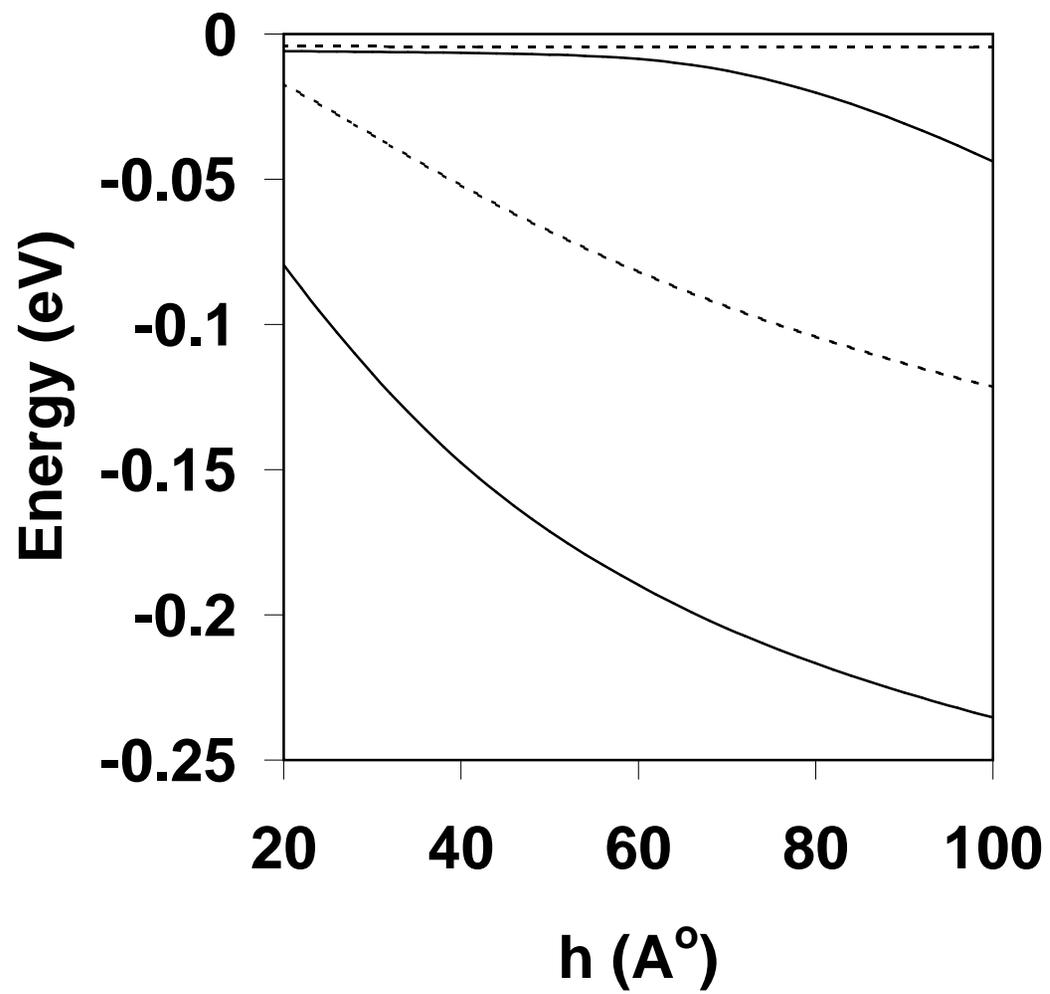
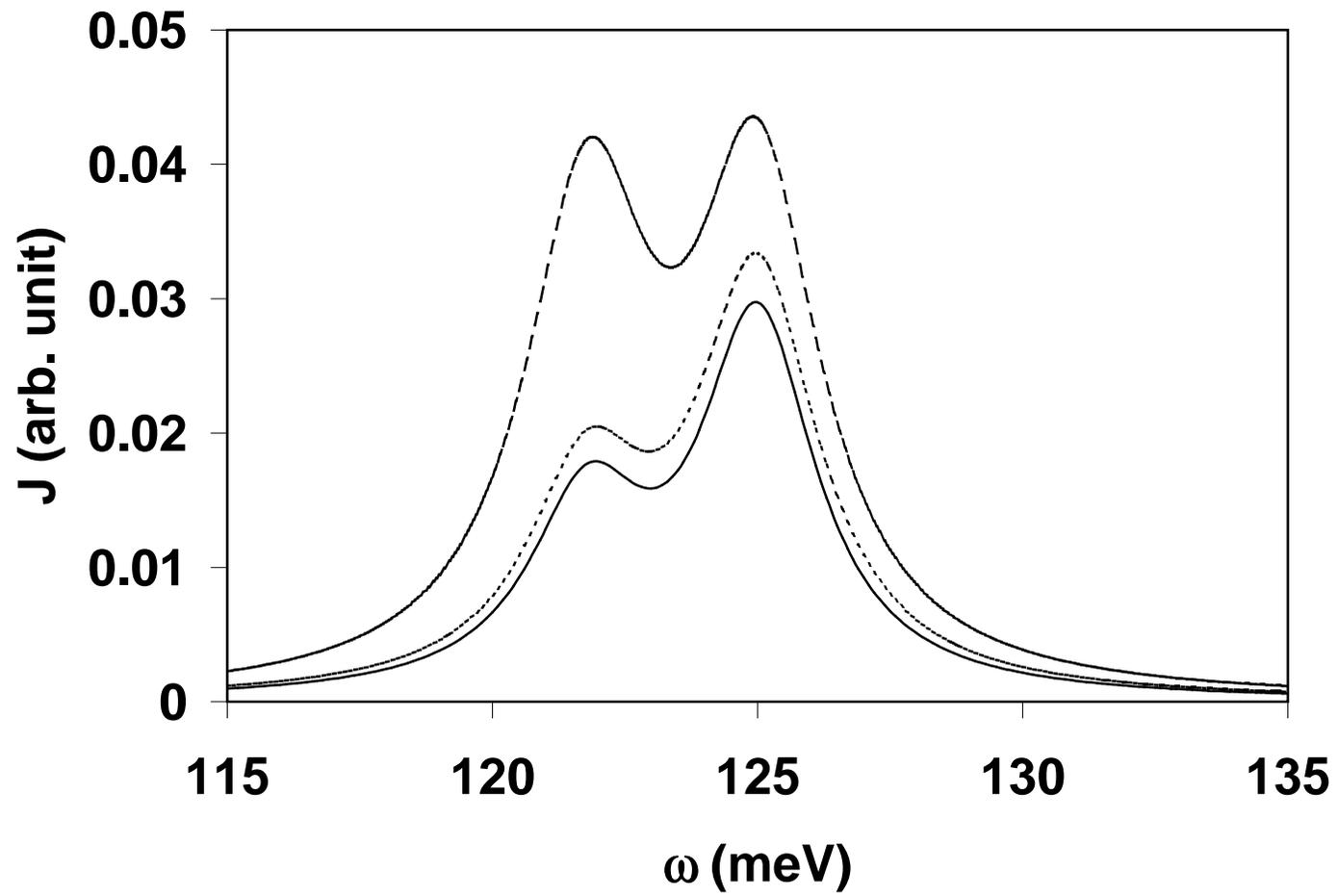
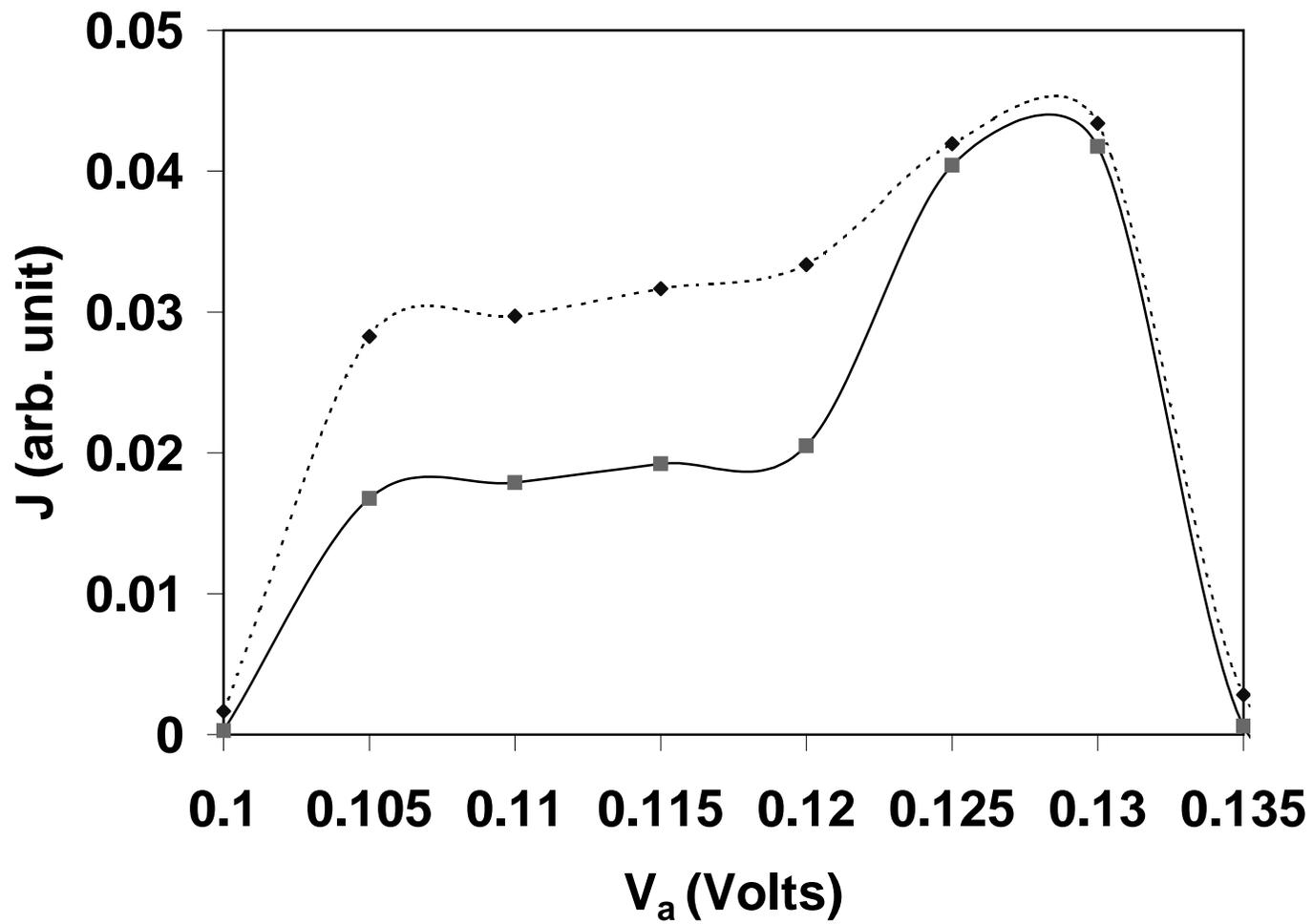
\end{document}